# Nanopore-patterned CuSe drives the realization of PbSe-CuSe lateral heterostructure

Bo Li,[1] Jing Wang,[1] Qilong Wu,[1] Qiwei Tian,[1] Ping Li,[2,*] Li Zhang,[1] Long-Jing Yin,[1] Yuan Tian,[1] Ping Kwan Johnny Wong,[3] Zhihui Qin,[1,*] and Lijie Zhang[1,*]

[1] Key Laboratory for Micro/Nano Optoelectronic Devices of Ministry of Education & Hunan Provincial Key Laboratory of Low-Dimensional Structural Physics and Devices, School of Physics and Electronics, Hunan University, Changsha 410082, People's Republic of China

[2] State Key Laboratory for Mechanical Behavior of Materials, Center for Spintronics and Quantum Systems, School of Materials Science and Engineering, Xi'an Jiaotong University, Xi'an, Shaanxi 710049, People's Republic of China

[3.] School of Microelectronics, Northwestern Polytechnical University, Xi'an 710072, Shaanxi & NPU Chongqing Technology Innovation Center, Chongqing 400000, People's Republic of China

Email: pli@xjtu.edu.cn; zhqin@hnu.edu.cn; lijiezhang@hnu.edu.cn

**Abstract**

Monolayer PbSe has been predicted to be a two-dimensional (2D) topological crystalline insulator (TCI) with crystalline symmetry-protected Dirac-cone-like edge states. Recently, few-layered epitaxial PbSe has been grown on the $SrTiO_3$ substrate successfully, but the corresponding signature of the TCI was only observed for films not thinner than seven monolayers, largely due to interfacial strain. Here, we demonstrate a two-step method based on molecular beam epitaxy for the growth of the PbSe-CuSe lateral heterostructure on the Cu(111) substrate, in which we observe a nanopore patterned CuSe layer that acts as the template for lateral epitaxial growth of PbSe. This further results in a monolayer PbSe/CuSe lateral heterostructure with an atomically sharp interface. Scanning tunneling microscopy and spectroscopy measurements reveal a four-fold symmetric square lattice of such monolayer PbSe with a quasi-particle band gap of 1.8 eV, a value highly comparable with the theoretical value of freestanding PbSe. The weak monolayer-substrate interaction is further supported by both density functional theory (DFT) and projected crystal orbital Hamilton population, with the former predicting the monolayer's anti-bond state to reside below the Fermi level. Our work demonstrates a practical strategy to fabricate a high-quality in-plane heterostructure, involving a monolayer TCI, which is viable for further exploration of the topology-derived quantum physics and phenomena in the monolayer limit.

**Introduction**

To date, various kinds of monolayer two-dimensional (2D) materials, such as graphene and transition metal dichalcogenides (TMDs), can be physically or chemically isolated from their parent crystals. Being in the monolayer limit, these 2D materials often exhibit richer properties that are not found in their 3D forms.[1] The 2D topological insulating phase is one such exotic state of matter known for elemental 2D materials such as silicene,[2-7] germanene,[8-13] stanene[14-16] and bismuthene,[17] in which the theoretically predicted quantum spin Hall effect may be explored at experimentally accessible temperatures.[18-21] In addition, substrate effect is of crucial to the topological properties.[22] On the other hand, the synthesis of binary honeycomb structure of $Sn_2Bi$ provides a platform for the studies of strongly correlated phenomena.[23] Along with other members of group IV-VI components that are best known for applications in thermoelectrics, optoelectronics, and spintronics,[24-27] PbSe is a narrow gap semiconductor crystallized in a rock salt structure. Remarkably, recent theory has predicted PbSe to possess a 2D topological crystalline insulator (TCI) phase when its physical dimension is reduced to monolayer.[28] While a vigorous proof of such a 2D TCI phase is yet to be seen, possible signature has been reported for epitaxial PbSe with seven layers grown on $SrTiO_3$ substrate. Lowering the layer number further for the search of the 2D TCI is, however, limited by interface strain that breaks the crystal symmetry of PbSe. To this end, achieving strain-free monolayer PbSe is vital.

Here, we demonstrate the synthesis of quasi-freestanding monolayer PbSe on Cu(111) by means of two-step molecular beam epitaxy, in which a nanopore patterned CuSe layer has been employed as the template for lateral epitaxy of PbSe, thereby forming a monolayer PbSe/CuSe lateral heterostructure. Scanning tunneling microscopy (STM) measurements show that the synthesized PbSe exhibits a four-fold symmetry with a periodicity of 0.43 nm. A quasi-particle band gap of 1.8 eV has been obtained by scanning tunneling spectroscopy (STS), revealing no major hybridization between monolayer PbSe and Cu(111) substrate. Our experimental results are in line with density functional theory (DFT) and projected crystal orbital Hamilton population (pCOHP), confirming a band gap of ~1.5 eV for freestanding PbSe and weak monolayer-substrate interaction.

**Results and Discussion**

Figure 1a shows a typical STM topographic image of large-scale Cu(111) substrate. After the deposition of a sub-monolayer (ML) of Se atoms and subsequent annealing at ~650 K for 10 min, monolayer CuSe decorated with patterned nanopores is formed (Fig. 1b). Previous reports indicated that the nanopores are related to Se-concentration and originated from the lattice mismatch between CuSe and Cu(111) substrate.[29-31] In accordance with the previous findings,[29] these triangular nanopores exhibit a hexagonal structure with a periodicity of ~3 nm and a depth of ~70 pm, as confirmed by the line profile in Figure 1b. Figure 1c shows a fast Fourier transform (FFT) of Figure 1b, revealing a hexagonal periodicity of the nanopores. It is worth mentioning that the nanopores exhibit not only triangular but also parallelogram shapes. The morphology of CuSe shown in Supporting Information Figure S1 further reveals the zigzag edges of the nanopores. To form PbSe, Pb atoms were evaporated to the surface in ultrahigh vacuum. Subsequent annealing at ~650 K for 10 min leads to the formation of a thin film island (Figure 1d). It is seen that the island possesses an apparent height of only 70 pm, much less than one atomic layer thickness of PbSe. This height difference strongly suggests the formation of PbSe atop Cu(111), instead of directly on CuSe, thus forming a lateral heterostructure with CuSe, which is schematically depicted in Figure S2 in the Supporting Information. We will elaborate this in more detail in latter section. X-ray photoelectron spectroscopy (XPS) was conducted to verify the valence states of Pb and Se ions. Figure 1e shows the main peaks at 137.9 and 142.9 eV corresponding to $Pb4f_{7/2}$ and $Pb4f_{5/2}$, respectively, of PbSe.[32] Similarly, Figure 1f indicates the two Se3d peaks at 54.05 and 54.7 eV, due to PbSe and CuSe.

To gain more information of the surface, we investigated the structural and electronic properties of monolayer PbSe and the surrounding CuSe. Figure 2a shows an atomic resolution STM image of PbSe. Clearly, the PbSe exhibits a four-fold symmetric square lattice, with its FFT shown in the inset of Figure 2a. The proposed top- and side-view of the PbSe ball-stick model is shown in Figure 2b. In order to determine the structure of PbSe on Cu(111), we investigated the thermodynamic stability of single atom layer (space group *P4/mmm*) and double atom layer (space group *P4/nmm*) PbSe, as shown in Figure S3. For the *P4/mmm* PbSe, it has relatively large imaginary frequency, implying its low stability. While for the *P4/nmm* PbSe its imaginary frequency almost disappears. Note that the small imaginary frequencies near the Γ point are probably due to the numerical errors. Hence, we consider the *P4/nmm* PbSe to be dynamically stable. As such, it is the *P4/nmm*, rather than *P4/mmm*, PbSe monolayer, that formed on the Cu(111) substrate, with two sublayers.[33] In Figure 2a, we obtain a structural periodicity of 0.43 nm and thus the lattice constant of the synthesized monolayer PbSe. In order to know which atoms are resolved by STM, we simulated the STM image, as shown in Figure S4, and found that only Pb atoms were visible under the empty state, which might be due to the higher surface height of Pb than Se atoms. This finding is similar to the cases of PbTe and SnTe, where only Pb and Sn atoms can be resolved by STM.[34-35] In a previous study of few-layered PbSe grown on $SrTiO_3$, the authors observed considerable compressive strain due to the mismatch between PbSe and the substrate.[36] Here, on the contrary, we have not found any strain effect, even though PbSe and Cu(111) are not structurally matched. For instance, the structure of Cu(111) is hexagonal with a three-fold symmetry, whereas PbSe exhibits a four-fold symmetric square structure. The incommensurate epitaxy involved here reduces the PbSe-substrate interaction, which also explains the absence of strain-induced lattice constant variation in our monolayer PbSe film. The resulting quasi-freestanding nature of the monolayer has also been verified by the *dI/dV* spectrum in Figure 2c. One can see that the valance band maximum (VBM) and conductance band minimum (CBM) are located at -0.65 eV and 1.15 eV, respectively, thus giving a quasi-particle band gap of 1.8 eV. Our measured large band gap reveals that no hybridization occurs between PbSe and Cu(111) substrate.[37] An atomic resolution STM image of CuSe around PbSe is shown in Figure 2e. The parallelogram nanopores distribute on the surface with varied orientations. A ball and stick model has been shown in Figure 2f, constructed based on the observation of CuSe structure, in accordance with previous report.[29] A line profile in Figure 2h along the line marked in Figure 2e shows the periodicity of the CuSe is ~0.41 nm, fitting quite well with the previous report.[29] The differential conductance spectrum in Figure 2g further indicates a metallic character of the CuSe layer. To gain more information of the formation mechanism of PbSe, we have scrutinized the growth procedure. We increased the concentration of Se to more than 1 ML on Cu(111) to obtain CuSe without nanopores pattern.[30-31] Figure 3a shows a large-scale STM image of nanopores-free CuSe. A zoom-in atomic resolution STM image shown in Fig. 3b reveals the same lattice parameters (see the inset in Figure 3b of ~0.41 nm) of nanopores-free CuSe and patterned one). Figure 3c shows an STM image of the surface after the deposition of Pb atoms. The marked bright protrusions are Pb clusters prior to annealing. Upon annealing at ~650 K for 10 min, the adsorbed Pb clusters disperse to Pb adatoms rather than forming PbSe/CuSe heterostructure (see in Figure 3d). The failed formation of PbSe suggests the crucial role of the nanopore patterned CuSe. Nevertheless, as aforementioned, the nanopore patterned CuSe could be realized by deposition of sub-ML Se on Cu(111), as also shown in Figure 3e. As mentioned before, there are two types of the nanopores of the patterned CuSe. They are visible in the two regions in Figure 3e labeled as I and II to distinguish them. Afterward, we deposited Pb atoms on patterned CuSe surface. Obviously bright protrusions of Pb adatoms/clusters are visible, distributed randomly on nanopore patterned CuSe surface as shown in Figure 3f. Subsequent annealing leads to the co-existence of PbSe and CuSe (see Figure 3g). Figure 3h and i show the formed PbSe/CuSe in-plane heterostructure with two types of interfaces labeled as I' and II', which correspond to the type I and II nanopores. Figure 3j and k illustrate the proposed schematic model for the formation mechanism of PbSe/CuSe lateral heterostructure. Two types of intrinsic nanopores decorated CuSe are schematically shown in Figure 3j, corresponding to that in Figure 3e. The nanopores expose dangling bonds that form local nucleation sites for PbSe, whereas the nanopore-free terraces of CuSe are fully saturated, thus preventing the nucleation. Furthermore, the interaction of the substrate might also take a role of preventing the nucleation. We proposed a mechanism of lateral heteroepitaxy PbSe templated by patterned CuSe as shown in Figure 3k. The Se atoms side of CuSe are bonded with Pb to achieve an epitaxy with PbSe and the growth is boosted to the other three directions (see the marked arrows in Figure 3k). As such, we explicitly conclude that the formation of PbSe/CuSe lateral heterostructure is related to nanopore patterned CuSe.

In order to further confirm the formation mechanism of PbSe/CuSe lateral heterostructure, we focus on the structural and electronic properties of the interface. Figure 4a shows a large-scale STM image of the monolayer PbSe and CuSe on the Cu(111) substrate. Figure 4b shows an atomic resolution STM image of simultaneously resolved with atomic resolution of PbSe and CuSe. Clearly, PbSe exhibits a square lattice whereas CuSe a honeycomb one. A zoom-in STM image of the interface is shown in Figure 4c, indicating the two sides are connected with each other by means of an atomic sharp interface, indicating the lateral heteroepitaxial growth of PbSe based on CuSe. It is comparable to the previous report of hBN and graphene,[38] whereas the difference between the two cases is that the requirement of consummation of nanopores-patterned CuSe to grown PbSe. As the proposed schematic aforementioned in Figure 3k, we find the interface boundary is consisted of Se bond with Pb atoms (see also the highlight ball and stick model inset of Figure 4c). It indicates that our proposal model fit quite well with the experimental observations, revealing the lateral heterostructure formed this way. It also explains the measured height of the PbSe islands of only ~70 pm (see Figure 4d), far from one atom thick. In addition, the interface interaction between PbSe and Cu(111) is weak as discussed before. Lateral heterostructure based on 2D materials have received extensive attention in recent years.[39-40] Among them, graphene-hBN and TMDs-based lateral heterostructures are the most extensively studied ones.[41-42] Heteroepitaxial growth of hBN templated by graphene edges opened pathways of construction hBN-graphene lateral heterostructure with atomic clean lateral interface.[38] Recently, these lateral heterostructure has expanded to other systems, for instance borophene-graphene and borophene-hBN lateral heterostructures.[43-44] Here in our case, we expand the strategy further. The nucleation occurs inside the nanopores of the patterned CuSe. By templating the nanopores patterned CuSe, PbSe grows laterally follow the marked other three directions (see Figure 3k) by dissociation the CuSe. Therefore, PbSe islands would forms and surrounds by CuSe, which is in line with our experimental observations. We calculated the formation energy by $E_f=(E_{tot} - N_{Cu}\mu_{Cu}/N_{Pb}\mu_{Pb} - N_{Se}\mu_{Se})/N_{tot}$, obtaining the formation energy of CuSe and PbSe are -3.48 eV and -4.20 eV, respectively. It is noted that the formation energy of PbSe is lower than CuSe, revealing the easier formation of PbSe. We conducted the differential conductance follow the line marked in the inset of Figure 4b. The series *dI/dV* spectra verify the transition from semiconducting to metallic properties from PbSe to CuSe. Room temperature induced noisy of the *dI/dV* spectra mainly in high energy zone, which would not influence the characteristics in the vicinity of the Fermi level. Figure 4f shows a color mapping of the real space based on the series *dI/dV* spectra spatially obtained from the surface, where we clearly find the band structure transition of PbSe to CuSe. We continue deposited Pb on PbSe to check its absorption, finding that the Pb atoms are absorbed on the edges of PbSe islands rather than on top of PbSe (see Figure S4 in Supporting Information). Subsequent annealing leads the expand of monolayer PbSe islands rather than forming the second layer of PbSe. Therefore, we further verify the monolayer feature of PbSe.

To support our observation in experiments, we performed the first principal calculations based on DFT. We have performed an extensive search for the stable structure of CuSe/Cu(111) and PbSe/Cu(111). We notice that the $\sqrt{3}\times\sqrt{3}$ ($\sqrt{10}\times1$) unit cell of CuSe (PbSe) is commensurate to the $\sqrt{7}\times\sqrt{7}$ ($2\sqrt{7}\times\sqrt{3}$) Cu(111) with a lattice mismatch of less than 5%. We have considered three typical configurations (see Figure S5 in Supporting Information), the Pb/Cu atom being in the top, bridge, and hollow positions of the quadrilateral formed by the top layer Cu atoms, respectively.

After extensive geometry optimizations, we found that the bridge CuSe/Cu(111) and hollow PbSe/Cu(111) structure are the most stable structure. For CuSe/Cu(111) heterostructure, top structure is more consistent with the experimentally grown structure. Hence, we take top CuSe/Cu(111) and hollow PbSe/Cu(111) as the objects for the following investigation.

To quantify the interaction between CuSe/PbSe and Cu(111) substrate, we calculated the binding energy $E_b$, defined as $E_b = (E_{CuSe/PbSe}+E_{Cu(111)}-E_{tot})/N$. $E_{CuSe/PbSe}$, $E_{Cu(111)}$ and $E_{tot}$ are the total energies of the CuSe/PbSe, the Cu(111) substrate, and the total energies of the CuSe/Cu(111) or PbSe/Cu(111). N is the number of the CuSe/PbSe atoms at the interface. The calculated $E_b$ is shown in Table I. The binding energy of CuSe is almost twice that of PbSe. It indicates that the interfacial interaction between PbSe and Cu(111) is much weaker than that between CuSe and Cu(111). In addition, the binding energies (0.61-0.62 eV/atom) are comparable to that of CrI$_3$ on semiconductor (0.28-0.41 eV/Cr)[45] and BiFeO$_3$ (0.34-0.51 eV/I)[46] substrate, indicating the nature of vdW interaction between PbSe and Cu(111).

Figure 5 illustrates the total density of states (TDOS) of CuSe and PbSe with/without the Cu(111) substrate. We find that CuSe film is metallic no matter the substrate stays there or not. However, the situation of PbSe case is completely different. When the substrate is there, the heterostructure shows metallic because the electronic state of the metal Cu(111) substrate is included, whereas it shows semiconducting without the substrate. Figure 5b shows the TDOS of freestanding PbSe without Cu(111) substrate by HSE06 functional, showing a band gap of ~1.5 eV, which is, actually, in accordance with our experimental measurements. It indicates that the weak interaction between PbSe and Cu(111) substrate, which is consistent with our experimental observations.

In order to systematically understand the physical mechanism, the synthesized semiconducting monolayer PbSe is weakly coupled from the substrate, we identify chemically bonded atom pairs via electron localization function (ELF) analysis. It can be a good substitute for the ambiguous distance cutoff method commonly used in previous work[47]. It is defined as ELF = $1/[1+(D/D_h)]$, where $D = \frac{1}{2}\sum_i |\nabla \varphi_i|^2 - \frac{1}{8}\frac{|\nabla \rho|^2}{\rho}$ and $D_h = \frac{3}{10}(3\pi^2 \rho)^{5/3}$. The $\varphi_i$ denotes the Kohn-Sham orbitals and $\rho = \sum_i |\varphi_i|^2$ represents electron charge density. The ELF provides a value between 0 and 1. ELF = 0.5 stands for the same level of Pauli repulsion as in the homogeneous electron gas, while a higher ELF value represents that the electrons are more localized (ELF = 1 indicates perfect localization of electrons). Figure S6 in SI shows the ELF for CuSe/Cu(111) and PbSe/Cu(111), which is a typical nearly free electron states at the interface. We found that the value of ELF between CuSe, PbSe layer and the Cu(111) substrate are lower than 0.5, indicating a weak interfacial interaction. In order to quantify and compare the strength of the interface interaction between CuSe and Cu(111), PbSe and Cu(111), we introduced pCOHP to analyze the interface interaction. We follow the usual way of displaying -pCOHP, namely, the positive values of -pCOHP, indicating the bonding states while negative values denote the antibonding states. For CuSe/Cu(111) (Figure 6a) and PbSe/Cu(111) (Figure 6b), the interface has antibonding states within E≈$E_F$-2.5 eV, and E≈$E_F$-2.0 eV (in the valence bands), indicating that the interface interaction is weaker. In order to better demonstrate the COHP results, the ipCOHP is defined as the following: -ipCOHP=$\int_{-\infty}^{E_F}$ -pCOHPdE, -ipCOHP indicates to what extent the interface interaction strength. The value of -ipCOHP is smaller, indicating the weaker interface interaction. The -ipCOHP values for CuSe/Cu(111) and PbSe/Cu(111) are 5.06 and 1.54 respectively. It indicate that the interface interaction of the PbSe/Cu(111) is much weaker than that of CuSe/Cu(111). It well explains why the *dI/dV* spectrum of PbSe/Cu(111) experimentally shows semiconducting property. Moreover, in order to confirm the weak van der Waals between PbSe and Cu(111) substrate, we calculated and compared the effect of the substrate on the band structures. As shown in Figure S7, we found that the Cu(111) substrate has little effect on the band structure. It indicate that there is a weak van der Waals interaction between PbSe and Cu(111) substrate, which can be ignored.

**Conclusions**

In summary, we have successfully synthesized monolayer PbSe by sequential deposition of Se and Pb on Cu(111). We find that nanopore patterned CuSe acts as a template for lateral epitaxy of PbSe, forming PbSe/CuSe lateral heterostructure. STM and STS shows the structural and electronic properties of PbSe, confirming its four-fold symmetry directly locate on Cu(111) with a weak interaction. The detailed atomic lattice of the obtained PbSe shows a lattice constant of 0.43 nm, revealing no strain effect exist. Combining STM/STS and DFT calculations, we find the obtained monolayer PbSe is in weak interaction with the substrate. Our work expands the strategy of lateral epitaxy of heterostructures based on 2D materials, paving the way toward the production of high-quality monolayers of 2D materials for both fundamental studies and potential applications.

**Experimental and Calculation Details**

*Experimental Details*

Our experiments were carried out in a home-built ultrahigh vacuum molecular beam epitaxy (UHV-MBE) chamber equipped with an STM (Unisoku). The base pressure of the system is better than $1\times10^{-10}$ Torr. The monolayer PbSe was epitaxially grown on Cu(111) substrate by the two-step method in the MBE chamber. Prior to the growth, the Cu(111) substrate was cleaned by several circles of sputtering (800 eV) and annealing (~850 K, 10 min). Subsequently, Se atoms (99.999%, Alfa Aesar) were thermally evaporated onto the Cu(111) substrate held at room temperature. Then, molecular beam Pb atoms were deposited on the surface. All the STM/STS measurements were conducted at room temperature. The STS (*dI/dV-V* curve) measurements were acquired by using a standard lock-in technique (793 Hz, 40-50 mV a.c. bias modulation). The system was carefully calibrated by the Si(111)-(7×7) and Au(111) surface. The characterization of the sample was also performed with XPS (Kratos Analytical Ltd. AXIS SUPRA with monochromatic 150w Al Ka X-ray).

*Calculation Details*

First, we used the Device Studio to build the crystal structures involved (Hongzhiwei Technology, Device Studio, Version 2021A, China, 2021. Available online: *https://iresearch.net.cn/cloudSoftware*). Then, we employed the Vienna ab-initio Simulation Package (VASP)[48] for the first-principles calculations based on DFT. A generalized gradient approximation (GGA) in the form of Perdew–Burke–Ernzerhof functional and HSE06 hybrid functional[49] was adopted for the exchange-correlation functional.[50] The van der Waals (vdW) density functional (optB86b-vdW) was capable of treating the dispersion force.[51] The energy convergence value between two consecutive steps was chosen as $10^{-5}$ eV and maximum force of 0.01 eV/Å was allowed on each atom. The plane-wave basis set with a kinetic energy cutoff of 400 eV was employed. The Cu(111) substrate was simulated by a repeating slab model consisting of a three Cu(111) slab with the calculated lattice constant of 2.56 Å. A 20 Å vacuum slab was considered to avoid the interaction between the supercell with its image. Γ-centered Monkhorst−Pack mesh was used to sample the Brillouin zone of the supercells.[52] To investigate the bonding situation of the interface, we performed a comprehensive bond analysis using LOBSTER package.[53] LOBSTER allows to project the plane-wave functions to the precious bonding information.

**Supporting Information**

The Supporting Information is available free of charge at **https://pubs.acs.org/**.

> The Supporting Information provides additional of experimental and calculated results (Figures S1-S6) (PDF).

**Conflict of interest**

> The authors declare that they have no conflict of interest.

**Acknowledgments**

This work was supported by the National Natural Science Foundation of China (Grant Nos. 11904094, 51972106, 12174096, 12004295 and 12174095), the Strategic Priority Research Program of Chinese Academy of Sciences (Grant No. XDB30000000), the Natural Science Foundation of Hunan Province, (Grant No: 2021JJ20026), the Natural Science Foundation of Shaanxi Province (Grant No: 2021JM-042), and the Natural Science Foundation of Chongqing, China (Grant No: cstc2021jcyj-msxmX0740). The authors also acknowledge the financial support from the Fundamental Research Funds for the Central Universities of China.

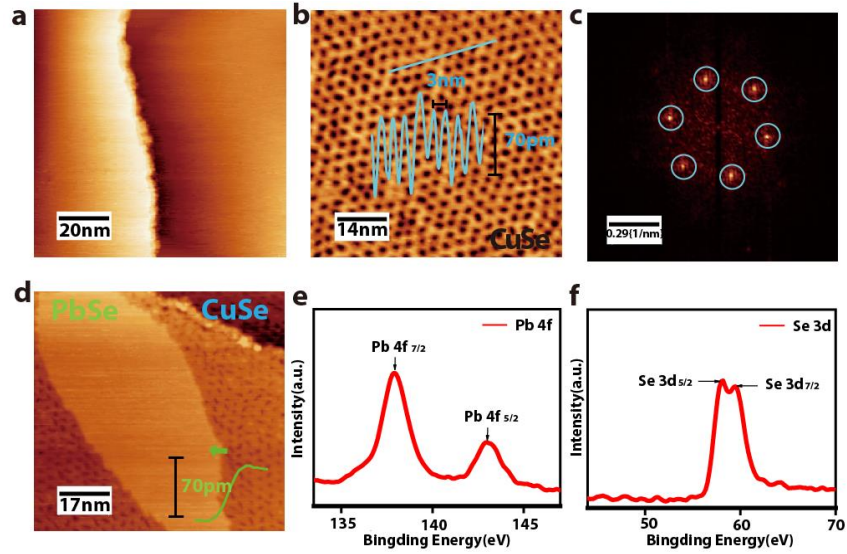

**Figure 1. Realization of monolayer PbSe.** (a) Large-scale STM image of clean Cu(111) substrate. Sample bias is 0.5 V and tunnel current is 80 pA. (b) Typical STM topographic image of CuSe with intrinsic triangular nanopores on Cu(111) substrate. An inset line profile reveals that the periodicity of nanopores pattern is ~3 nm, and the depth of the nanopores is ~70 pm(V=1.4 V, I=80 pA). (c) 2D Fast Fourier transform (FFT) pattern made from (b). The bright dots marked by cyan circles confirms the hexagonal periodicity of the nanopores. (d) STM topographic image of monolayer PbSe surrounded by CuSe. The inset height profile shows the step height of the film is ~70 pm(V=1 V, I=90 pA). (e, f) XPS spectra of the Se 3d and Pb 4f core levels, respectively.

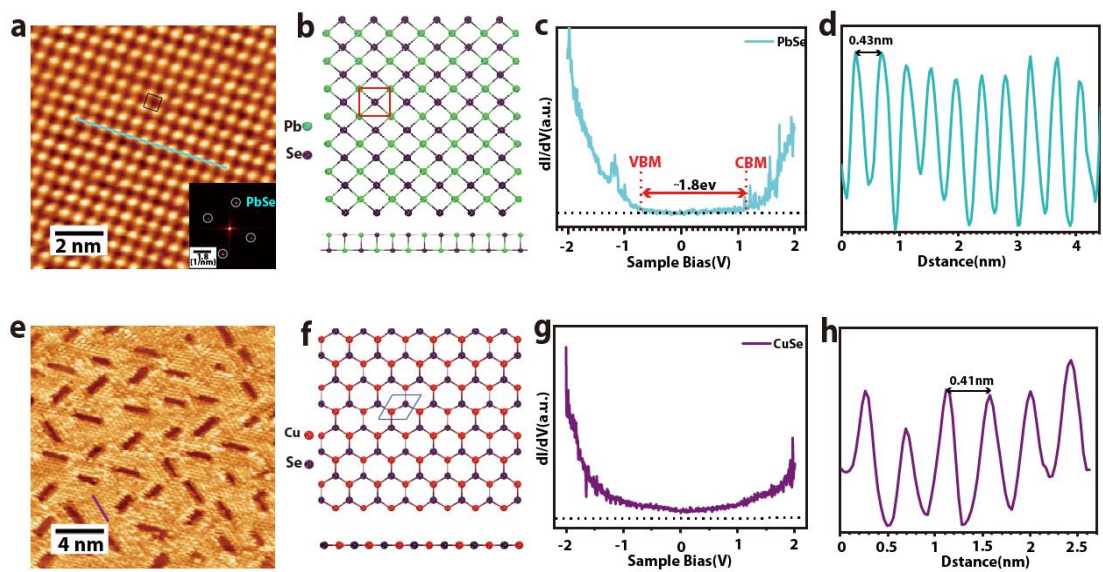

**Figure 2. Characteristic of PbSe and CuSe.** (a) High-resolution STM images of monolayer PbSe (V=-500 mV, I= 90 pA). Inset: the corresponding FFT pattern with the cyan circles marked the reciprocal lattice peaks of PbSe. (b) Ball and stick model of PbSe crystalline structure (upper panel: top view, lower panel: side view). (c) *dI/dV* spectra of monolayer PbSe, revealing a band gap of ~1.8 eV. (d) Line-profile of PbSe along the line marked in (a), revealing the periodicity of the synthesized PbSe is ~0.43 nm. (e) High-resolution STM images of monolayer CuSe (V=1 V, I=90 pA). (f) Ball and stick model of CuSe crystalline structure (upper panel: top view, lower panel: side view). (g) *dI/dV* spectra of monolayer CuSe, revealing a clear metallic property. (h) Line profile of CuSe along the line marked in (e), revealing the periodicity of CuSe is ~0.41 nm.

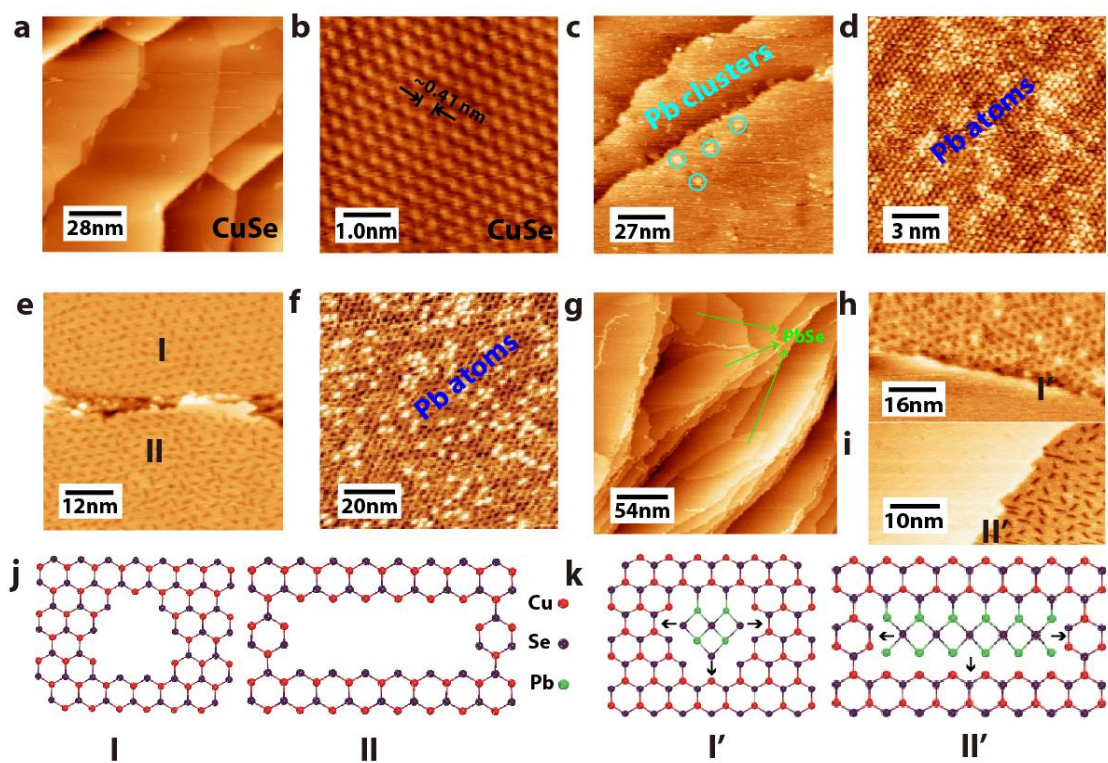

**Figure 3. Formation route of PbSe.** (a) Large-scale STM image of CuSe without nanopores (V= 2V, I= 10 pA). (b) Atomic resolution STM image of CuSe without nanopores on the Cu(111) substrate (V= 1 V, I=90 pA). (c) Large-scale STM topographic image of Pb evaporated on a nanopores-free CuSe (V= 1 V, I=100 pA). The marked bright protrusions are Pb clusters. (d) The STM topography after annealing of (b) (V= 1 V, I=90 pA). The bright dots are Pt atoms. (e) Large-scale STM image of the CuSe with two types of nanopores on the Cu(111) substrate (V=1.3 V, I=59 pA). Type I: triangle shape, and Type II: parallelogram shape. (f) STM topography of Pb evaporated on CuSe with nanopores (V= 1 V, I=90 pA). The bright protrusions are Pb clusters. (g) Large-scale STM topographic image of monolayer PbSe islands surrounding by CuSe. (V= 1V, I=90 pA). (h, i) STM images of PbSe/CuSe in-plane heterostructures based on two types of nanopores as shown in (e), labeled as I' and II', respectively. (V= 1V, I=90 pA) (g) STM topography of PbSe/CuSe lateral heterostructure obtained by annealing (e). (V= 700 mV, I=90 pA). (j) Ball and stick models for nanopores CuSe. Type I and II are triangle and parallelogram, respectively, corresponding to the STM image in (e). (k) Schematic diagram of the formation of PbSe/CuSe heterostructure. The left panel labeled by I' corresponds to I, whereas the right panel II' to II. The arrows are the growth direction of PbSe.

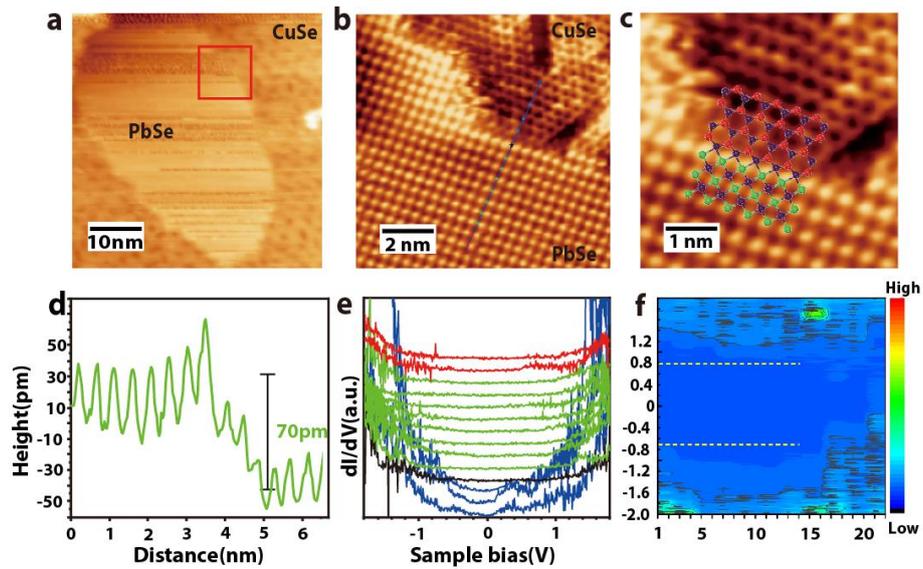

**Figure 4. Electronic properties of PbSe/CuSe lateral heterostructure.** (a) Large-scale STM image of the monolayer PbSe and CuSe on the Cu(111) substrate (V= 1.4V, I=40 pA) (b) Atomic resolution STM image of the monolayer PbSe and CuSe on the Cu(111) substrate (V= -500 mV, I=90 pA). (c) Zoom-in STM image of the interface of PbSe and CuSe. Ball and stick model on top to reveal the structural properties of the interface bounding. (d) A line profile along the green line marked in (b) shows the step height of ~70 pm. (e) Serials *dI/dV* spectra collected from PbSe to CuSe crosses the step edge, showing the electronic properties of PbSe and CuSe. (d) Color mapping of the real-space imaging of the band profile plotted in terms of the *dI/dV*.

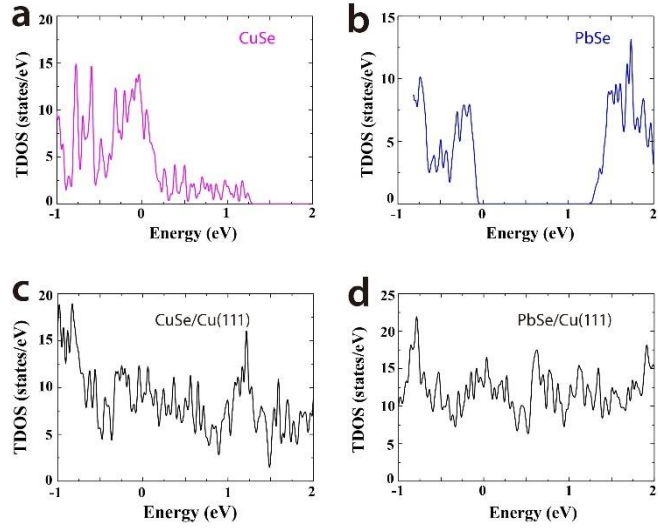

**Figure 5. Total density of states (TDOS) of CuSe and PbSe.** (a) TDOS of CuSe detached from Cu(111). (b) TDOS of PbSe detached from Cu(111) by HSE06 functional. (c, d) TDOS of CuSe/Cu(111) and PbSe/Cu(111), respectively. Zero refers to the Fermi level.

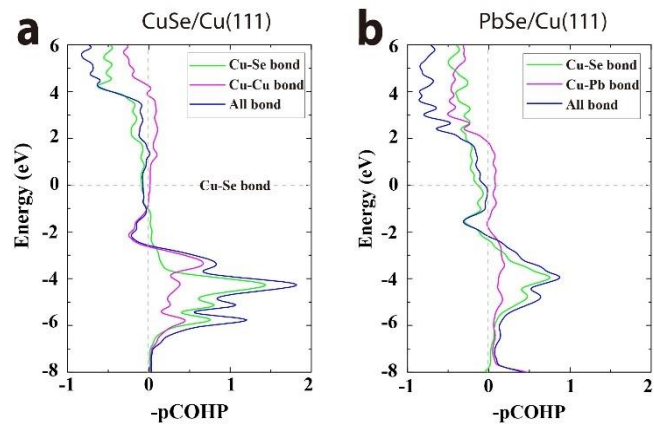

**Figure 6. Interaction comparison between the overlayer and the substrate.** Plane wave pCOHP between Cu and Cu, Se atoms (a) and between Cu and Pb, Se atoms (b) at the interface. -pCOHP > 0 denotes the bonding state, while -pCOHP < 0 indicates the antibonding state. Energy is shifted so that the Fermi level $E_F$ equals zero.

Table I. Binding Energy $E_b$ (eV/Atom) of the Three Structures for CuSe and PbSe on the Cu(111) Substrate.

|      | bridge (eV) | hollow (eV) | top (eV) |
|------|-------------|-------------|----------|
| CuSe | 1.12        | 1.12        | 0.87     |
| PbSe | 0.62        | 0.62        | 0.61     |